\documentstyle[preprint,eqsecnum,aps,epsf]{revtex}

\newcommand{\ksl}{k \! \! \! /}
\newcommand{\Omegasl}{\Omega \! \! \! \! /}
\newcommand{\Deltasl}{\Delta \! \! \! \! /}
\renewcommand{\theequation}{\arabic{equation}}

\begin{document}
\draft
\preprint{SAGA-HE-122-97 (hep-ph/9706420)}
\title{Two-loop anomalous dimensions for the structure function $h_1$}
\author{S. Kumano and M. Miyama\cite{byline}}
\address{Department of Physics, Saga University \\
         Saga 840, Japan}
\date{June 18, 1997}
\maketitle
\begin{abstract}
Chiral-odd structure function $h_1$ is expected to be measured 
in polarized Drell-Yan process.
We calculate two-loop anomalous dimensions for
$h_1$ in the minimal subtraction scheme. 
Dimensional regularization and Feynman gauge
are used for calculating the two-loop contributions.
Our results are important in studying $Q^2$ dependence of $h_1$.
\end{abstract}

\narrowtext


Internal spin structure of the proton has been a popular topic
since the discovery of a serious proton-spin problem at high
energies. In spite of a naive quark-model prediction, 
it was suggested that the major part of proton spin is not 
carried by quarks by measuring the structure function $g_1$.
The discovery inspired many theoretical and experimental
studies on the spin structure. In order to understand 
how the proton spin consists of quark and gluon spins,
it is important to learn about sea-quark and gluon
polarizations. As an experimental project to investigate
these issues, the RHIC (Relativistic Heavy Ion Collider) Spin project
was proposed.

In this project, the internal spin structure will be
investigated by polarized proton-proton reactions.
Another virtue of this project is that a chiral-odd
structure function $h_1$ could be measured in
the Drell-Yan process with transversely polarized protons 
\cite{trans,h1evol,jj,dyexp}.
It could be also measured in semi-inclusive electron scattering
\cite{h1ele}. It provides us another important test of proton spin
structure. In particular, the structure
function $h_1$ probes different aspects of spin.
Furthermore, a nice point is that it has a clear parton-model
interpretation. It measures the transversity distribution, namely
the probability to find a quark with spin polarized along
the transverse spin of a polarized proton minus
the probability to find it polarized oppositely.
Its gross properties have been studied recently \cite{h1cal}
in particular by Jaffe and Ji \cite{jj}.
Here, we would like to focus on $Q^2$ dependence.
Leading-order $Q^2$ evolution was already studied \cite{h1evol}.
However, it is our standard to use next-to-leading-order (NLO)
results in obtaining unpolarized and polarized parton distributions. 
Recently, the NLO evolution equations are established for
the structure function $g_1$ \cite{g1}.
Therefore, it is important to 
investigate the NLO $Q^2$ evolution of $h_1$ for future analyses
of polarized Drell-Yan data in connection with the
$h_1$ structure function.
Our research purpose is to calculate two-loop
anomalous dimensions, which determine the NLO $Q^2$ evolution
of $h_1$.

\begin{figure}
\vspace{2.0cm}
\hspace{0.0cm}
\epsfxsize=14.0cm
\centering{\epsfbox{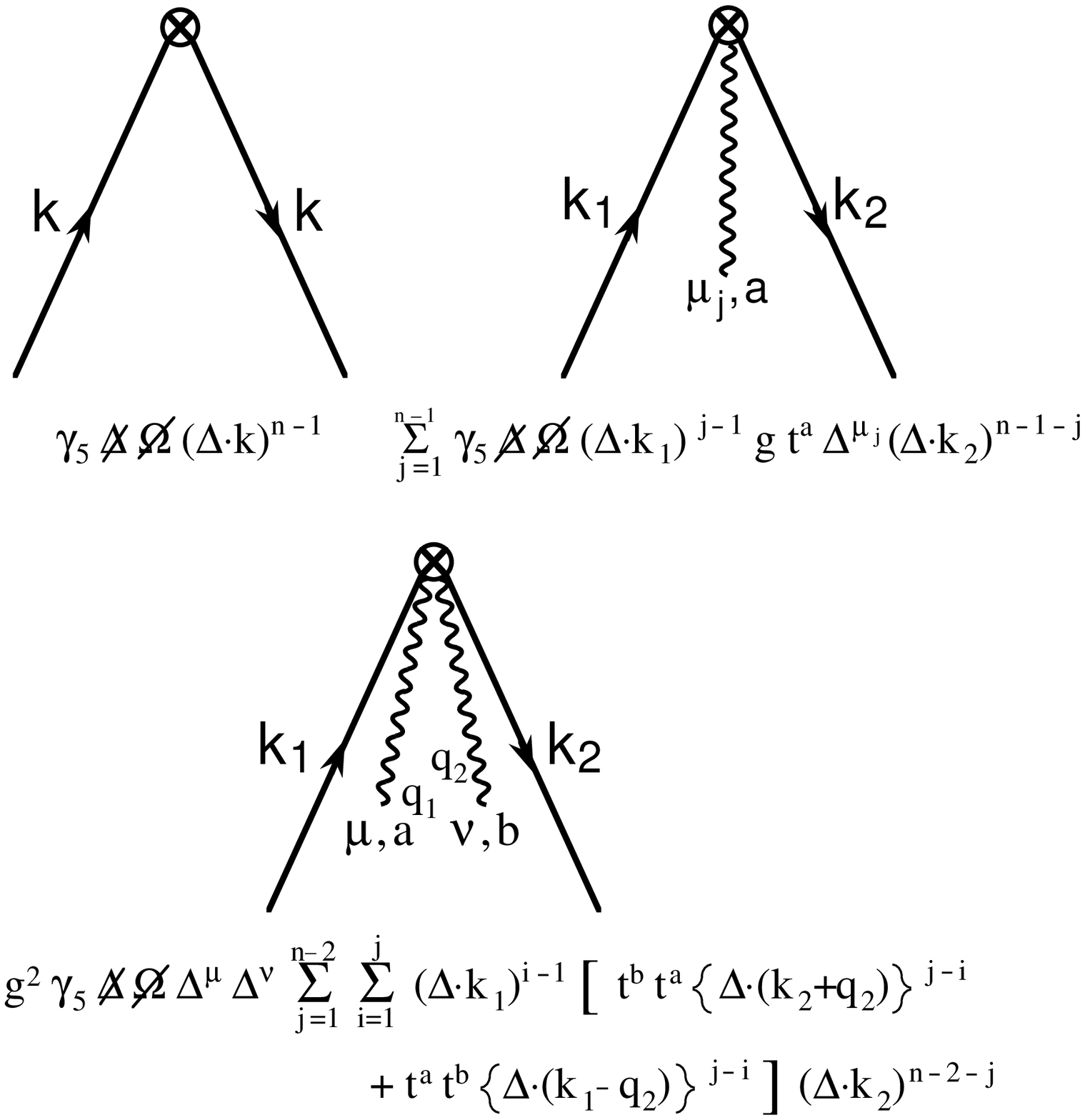}}
\vspace{1.5cm}
\caption{Feynman rules for zero, one, and two gluon vertices.}
\label{feyn}
\end{figure}

In order to study $h_1$ in perturbative Quantum Chromodynamics (QCD),
we need to introduce a set of local operators \cite{jj}
\begin{equation}
O^{\nu\mu_1 \cdot\cdot\cdot \mu_n}
   = S_n \, \overline\psi \, i \, \gamma_5 \, \sigma^{\nu\mu_1} \, 
         i D^{\mu_2} \cdot\cdot\cdot i D^{\mu_n} \, \psi
     - \ {\rm trace} \ {\rm terms}
\ , \ \ n=1, 2, ... 
\ \ \ .
\label{ope}
\end{equation}
Here, $S_n$ symmetrizes the Lorentz indices $\mu_1$, ..., $\mu_n$,
and $iD^\mu=i\partial^\mu + g t^a A_a^\mu$ is the covariant derivative.
We calculate two-loop anomalous dimensions of these twist-two operators.
Because gluon field cannot contribute to the anomalous dimensions of
the chiral-odd operators even in the NLO, our calculation procedure
becomes a rather simple one which is similar to the nonsinglet formalism 
for example by Floratos, Ross, and Sachrajda \cite{frs}.
Renormalization of the 
$\overline\psi \gamma^{\mu_1} (1\pm \gamma_5) 
i D^{\mu_2} \cdot\cdot\cdot i D^{\mu_n} \psi$
type operators was studied in Ref. \cite{frs}, but
the same method can be applied to the operators in Eq. (\ref{ope}).
First, Feynman rules at the operator vertices should be provided.
In order to satisfy the symmetrization condition and to remove
the trace terms, the tensor 
$\Delta_{\mu_1} \Delta_{\mu_2} \cdot\cdot\cdot \Delta_{\mu_n}$
with the constraint $\Delta^2=0$ is usually multiplied.
However, the operators in Eq. (\ref{ope}) are associated with one-more
Lorentz index $\nu$, so that it is convenient to introduce
another vector $\Omega_\nu$ with the constraint $\Omega\cdot\Delta$=0.
Then, Feynman rules with zero, one, and two gluon vertices become
those in Fig. \ref{feyn}.

Next, we give a brief outline how anomalous dimensions are calculated
\cite{frs}. The bare operator is defined by
$O_B^n =\overline\psi_B Q_B^n \psi_B =Z_O O_R^n$ with
the renormalized one $O_R^n$.
Then, the renormalization constant $Z_O$ is given by
$Z_O=Z_F/Z_Q$, where $Z_F$ and $Z_Q$ are 
wave-function and operator renormalization constants.
Anomalous dimensions for $O^n$, $Q^n$, and $\psi$ are
given by these renormalization constants as
$\gamma_{_{O^n}} = \mu \, \partial (\ln Z_{O^n})/\partial\mu$,
$\gamma_{_{Q^n}} = \mu \, \partial (\ln Z_{Q^n})/\partial\mu$, and
$\gamma_{_F}   = (\mu/2) \, \partial (\ln Z_{F})/\partial\mu$.
Using the relation among the renormalization constants,
we express the anomalous dimension $\gamma_{_{O^n}}$ 
in terms of $\gamma_{_{Q^n}}$ and $\gamma_{_F}$ as
$\gamma_{_{O^n}}=2\gamma_{_F}-\gamma_{_{Q^n}}$.
In dimensional regularization with the dimension $d=4-\epsilon$,
$Z_{Q^n}$ can be obtained in the form
$Z_{Q^n}=1+\sum_{k=1}^\infty Z_k^{Q^n} (g_{_R})/\epsilon^k$.
As it is shown in Ref. \cite{frs}, the anomalous dimension
is related to the first coefficient $Z_{k=1}^{Q^n}$ by
$\gamma_{_{Q^n}}=-g_{_R}^2 \, \partial Z_1^{Q^n}/\partial g_{_R}^2$.
In the two-loop calculation, it becomes
$\gamma_{_{Q^n}} = - \, 2 \, Z_1^{Q^n}$.
The anomalous dimension $\gamma_{_{Q^n}}$ is simply given
by minus twice the $1/\epsilon$ coefficient in $Z_{Q^n}$.
The same equation can be also applied to the anomalous
dimension $\gamma_{_F}$.
In this way, we try to find $1/\epsilon$ singularities
in the renormalization constants for
calculating the anomalous dimensions.

\begin{figure}
\vspace{0.0cm}
\hspace{0.0cm}
\epsfxsize=13.0cm
\centering{\epsfbox{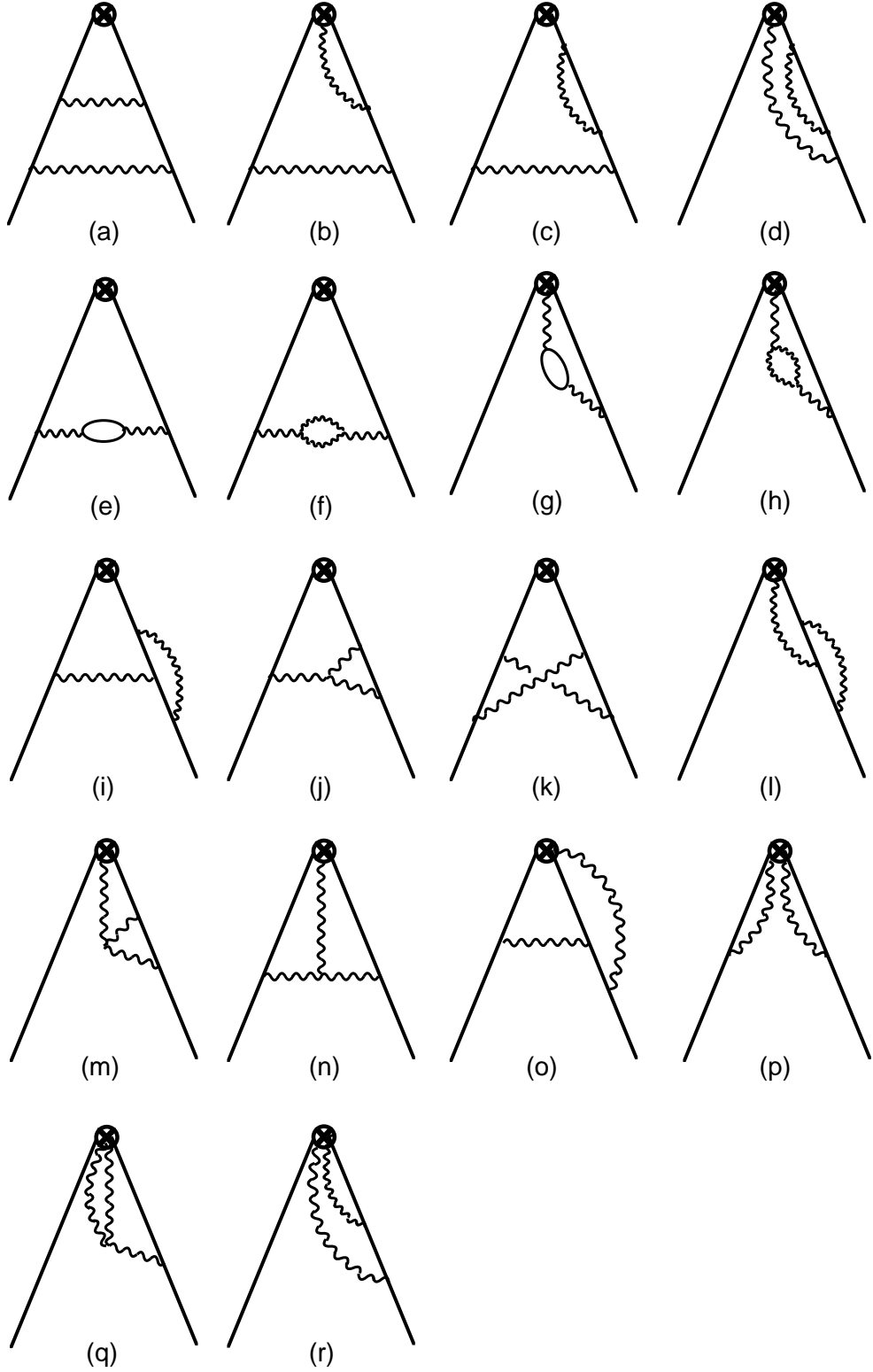}}
\vspace{1.2cm}
\caption{Contributions to the two-loop anomalous dimension.}
\label{diag}
\end{figure}

\begin{figure}
\vspace{0.0cm}
\hspace{0.0cm}
\epsfxsize=15.0cm
\centering{\epsfbox{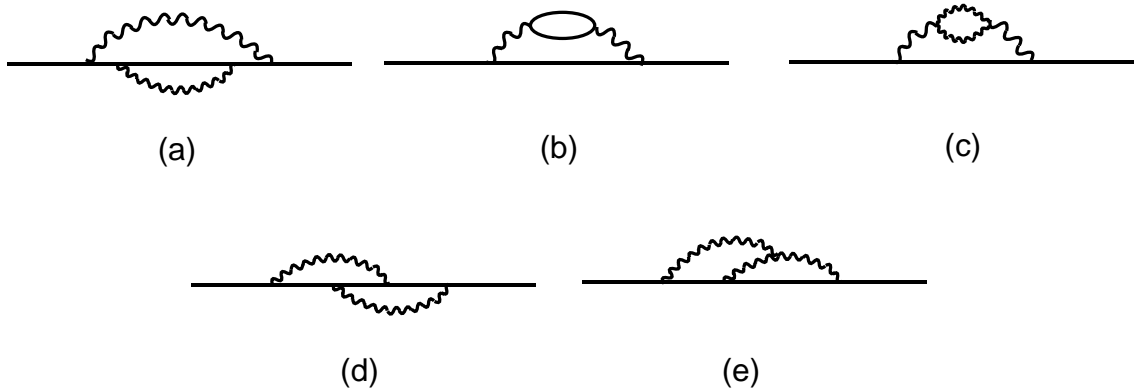}}
\vspace{1.5cm}
\caption{Two-loop contributions to the quark self-energy.}
\label{quark}
\end{figure}

\begin{figure}
\vspace{0.0cm}
\hspace{0.0cm}
\epsfxsize=8.0cm
\centering{\epsfbox{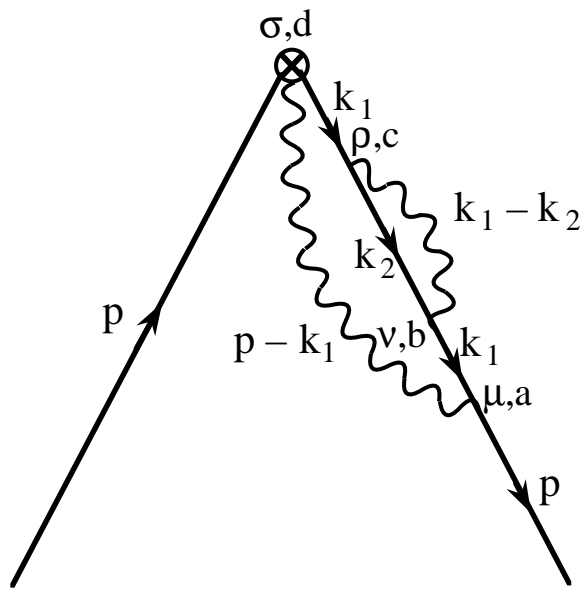}}
\vspace{1.5cm}
\caption{Lorentz and color indices are shown with momenta
         for the diagram in Fig. \ref{diag}(d).}
\label{d}
\end{figure}

Two-loop contributions are calculated by using
the dimensional regularization with the Feynman gauge.
Here, minimal subtraction (MS) scheme is used.
Possible Feynman diagrams contributing to the one-particle
irreducible matrix elements are shown in Fig. \ref{diag}.
Furthermore, there are two-loop diagrams contributing to the quark
self-energy as shown in Fig. \ref{quark}.
However, the quark-field renormalization is already
discussed in Ref. \cite{frs}, so that 
we do not have to repeat the same calculations.
In order to explain our calculation procedure for the operator part,
we pick a simple case.
In Fig. \ref{d}, Lorentz and color indices are shown with momenta
for the diagram (d) in Fig. \ref{diag}. 
The diagram is calculated with a Feynman rule in Fig. \ref{feyn} as
\begin{eqnarray}
I_d = \int \frac{d^dk_1}{(2\pi)^d} \frac{d^dk_2}{(2\pi)^d}
       \, ig t^a \gamma^\mu \, \frac{i \ksl_1}{k_1^2} 
       \, ig t^b \gamma^\nu \, \frac{i\ksl_2}{k_2^2} 
       \, \frac{-ig_{\nu\rho}\delta_{bc}}{(k_1-k_2)^2} 
       \, ig t^c \gamma^\rho \, \frac{i \ksl_1}{k_1^2} 
       \, \frac{-ig_{\mu\sigma}\delta_{ad}}{(p-k_1)^2}
\nonumber \\
        \times \, \, \sum_{j=1}^{n-1} \, \gamma_5 \Deltasl \, \Omegasl
       \, \, (\Delta\cdot p)^{j-1} 
       \, g t^d \Delta^\sigma \, (\Delta\cdot k_1)^{n-1-j}
\ \ \ .
\label{int-d}
\end{eqnarray}
Calculating the above integral, we obtain
\begin{eqnarray}
I_d=\gamma_5 \Deltasl \, \Omegasl \, \, (\Delta\cdot p)^{n-1}
    \, 2 (2-\epsilon) 
    \, g^4 C_F^2 \, \frac{1}{(4\pi)^{4}}
    \, \frac{\Gamma(\epsilon/2)\Gamma(\epsilon)}{\Gamma(1+\epsilon/2)}
    \, B(2-\epsilon/2,1-\epsilon/2) 
\nonumber \\
    \times \, \, \sum_{j=1}^{n-1} \,
        B(j+1-\epsilon,1-\epsilon/2) \, 
        \left ( \frac{-p^2}{4\pi} \right )^{-\epsilon}
\ \ \ ,
\end{eqnarray}
where $\Gamma (x)$ and $B(x,y)$ are gamma and beta functions,
and $C_F$ is given by
$C_F=(N_c^2-1)/(2N_c)$ with the number of color $N_c$.
Subtracting a one-loop counter term from the above equation
and evaluating singular terms, we have
\begin{equation}
I_d'= \gamma_5 \Deltasl \, \Omegasl \, \, (\Delta\cdot p)^{n-1}
      \, 2 \, \frac{g^4}{(4\pi)^4} \, C_F^2 \, 
      \sum_{j=1}^{n-1} \,
      \frac{1}{j+1} \left [ - \frac{2}{\epsilon^2}
         + \frac{1}{\epsilon} \, \{ 1-S_1 (j+1) \} \right ]
\ \ \ ,
\end{equation}
where $S_1(n)=\sum_{k=1}^n 1/k$.
From the $1/\epsilon$ singularity in the above equation,
we obtain the diagram (d) contribution to 
the anomalous dimension as
$\gamma_n^{(2d)} =  8 \, g^4 \, C_F^2 \, 
                    [ \, G_1(n) - S_1(n) \, ] /(4\pi)^4$
with $G_1(n)=\sum_{j=1}^n (1/j) \sum_{i=1}^j (1/i)$.
The factor of two is included by considering a similar diagram
with gluons attached to the initial quark line.
We find that the obtained anomalous dimension
is exactly the same with the one in Ref. \cite{frs}.
The reason is the following. 
Because the operator vertex part $\gamma_5 \Deltasl \, \Omegasl \, $
can be separated from the $k_1$ and $k_2$ integrals 
in Eq. (\ref{int-d}), the integrals are independent of the operator form.
The vertex $\Deltasl \,$ in Ref. \cite{frs} is simply replaced 
by the present one $\gamma_5 \Deltasl \, \Omegasl \, $.
Therefore, the renormalization calculations
are the same in both cases, so that the obtained anomalous
dimensions are exactly the same.

We notice that several diagrams in Fig. \ref{diag} have
the same property in the sense that the operator part
can be separated from the integrals.
These are cases with gluon propagators attached
only to the final-quark or initial-quark line.
In addition to the diagram (d), these are diagrams
(g), (h), (l), (m), (q), and (r). We do not have to repeat
the same calculations.
Furthermore, there are certain diagrams which do not contribute to
the anomalous dimension.
Calculating diagrams (a), (b), (c), and (p), we find
that there exist no $1/\epsilon$ type singularity.
Therefore, there is no contribution to the anomalous
dimension from these diagrams.
From the above discussions, the problem reduces
to calculations of remaining diagrams (e), (f), (i), (j),
(k), (n), and (o). 

We need to calculate integrals which appear in 
evaluating these diagrams.
Fortunately, many of them are already listed in Appendix of
Ref. \cite{frs}. 
All divergent subintegrals are renormalized by subtracting
the pole part of the subintegral. However, special care has to be
taken in some diagrams, for example the one in Fig. \ref{diag}(i).
Contracting $\gamma$ matrices in the numerator, we find
a term proportional to $\epsilon$. 
In this case, the $1/\epsilon$ contributions come from
the double-pole terms, in which it is known physically
that there is no counterterm contribution \cite{frs}. 
Therefore, when we write the double integral results,
it is more convenient to express the $1/\epsilon^2$ terms without 
subtracting counterterms even though the $1/\epsilon$ terms
are calculated with the counterterms. 
This is actually done in Appendix of Ref. \cite{frs}
for writing the integral results.
We also express extra integrals in this way in Appendix B.
Once we have all the integral information, it is straightforward
calculation for obtaining the anomalous dimensions.
Because calculation procedure is essentially 
the same with the one for the diagram (d)
and the details are already described in Ref. \cite{frs},
each evaluation is not explained. 
We simply list our results in Appendix A.

We summarize our study.
Two-loop anomalous dimensions for the chiral-odd structure
function $h_1$ are calculated in the MS scheme.
Dimensional regularization and Feynman gauge are used
for calculating the possible diagrams.
Our results are important for studying the details of
$Q^2$ dependence in $h_1$, which is expected to be measured
at RHIC.

Note added:
After we submitted this paper (hep-ph/9706420), two preprints
appeared in the Los Alamos archive on the same topic.
Our anomalous dimensions agree not only with the Vogelsang's
result (hep-ph/9706511) but also with the Hayashigaki, Kanazawa,
and Koike's (hep-ph/9707208) although the integral
expressions are different in some diagrams.

$\ \ \ $

\noindent
{\bf Appendix A: List of anomalous dimensions}

Contributions from the diagrams in Figs. \ref{diag} and \ref{quark}
to the anomalous dimension are listed.
In the following equations, $\hat H_k$ is defined by
$\hat H_k =(\Deltasl \, \Omegasl \, )^{-1}
           [H_k (n,b=0) \Omegasl \, - H_k (n-1,b=1) \Deltasl \, ]$,
and DP (SP) denotes double-pole (simple-pole) part of the integral.
The integrals $H_k$ are listed in Appendix B, and $I_k$ are
the integrals given in Appendix of Ref. \cite{frs}.
The function $\widetilde F$ is defined by
$\widetilde F(a,c) = \sum_{J=0}^ c
    (-1)^{a+J+1} \, c! \, F(a+J,c-J) /[J! \, (c-J)!]$.
Some results are exactly the same with those in Ref. \cite{frs}
as explained in the text:
\begin{eqnarray}
& & \gamma_n^{(2d)}, \ \gamma_n^{(2g)}, \ \gamma_n^{(2h)}, \ 
    \gamma_n^{(2l)}, \ \gamma_n^{(2m)}, \ \gamma_n^{(2q)}, \ 
    \gamma_n^{(2r)}, 
\nonumber \\
& & \gamma_n^{(3a)}, \ \gamma_n^{(3b)}, \ \gamma_n^{(3c)}, \ 
    \gamma_n^{(3d)}, \ \gamma_n^{(3e)}
\ \ \ \textstyle{\rm given \ in \ Ref. \cite{frs}}
\ \ \ .
\end{eqnarray}
Other anomalous dimensions are calculated as
\begin{equation}
\gamma_n^{(2a)} = \gamma_n^{(2b)} = \gamma_n^{(2c)} 
                = \gamma_n^{(2p)} = 0
\ \ \ ,
\end{equation}
\vspace{-1.0cm}
\begin{equation}
\gamma_n^{(2e)} = - \, \frac{g^4}{(4\pi)^4} \, C_F \, T_R \, \frac{16}{9}
\ \ , \ \ 
\gamma_n^{(2f)} = + \, \frac{g^4}{(4\pi)^4} \, C_F \, C_G \, \frac{32}{9}
\ \ ,
\end{equation}
\vspace{-1.0cm}
\begin{eqnarray}
\gamma_n^{(2i)} &=& + \, g^4 \, (C_F^2 - C_F C_G /2) \,
          4 \, (\Delta\cdot p)^{-(n-1)} \,  
   \bigg [ \, 4  \, \hat H_1^{DP} (n)
                  - 6 \, \hat H_2^{DP} (n) 
                  + 4 \, \hat H_3^{DP} (n) 
                  - 4 \, \hat H_{4}^{DP} (n) 
\nonumber \\
& &
          - 2 \, I_{2}^{DP} (0,n-1)
          -      I_{11}^{DP} (0,n-1) 
          + 2 \, I_{12}^{DP} (0,n-1)
          -   I_{13}^{DP} (0,n-1)
          + 3 \, I_{14}^{DP} (0,n-1)  
\nonumber \\
& &
     + (\Delta\cdot p)^{-1} \bigg \{ \,
              I_{2}^{DP}(0,n) - 2 \, I_{2}^{DP}(1,n-1) 
            - I_{12}^{DP}(0,n) +      I_{12}^{DP}(1,n-1) 
            + I_{13}^{DP}(0,n) 
\nonumber \\
& &   
             - I_{13}^{DP}(1,n-1) 
             - I_{14}^{DP}(0,n) + 2 \, I_{14}^{DP}(1,n-1)            
                                         \, \bigg \} \, 
            + 4 \, \hat H_2^{SP} (n)
            - 2 I_3^{SP} (n-1,0)
\nonumber \\
& & 
 + (\Delta\cdot p)^{-1} \bigg \{ \, 
             - 2 \, I_{1}^{SP}(0,n)   
             + 2 \, I_{2}^{SP}(0,n) 
             - 4 \, I_{2}^{SP}(1,n-1) 
             + 2 \, I_{3}^{SP}(n,0)  
\nonumber \\
& &            
             +  2 \, I_{3}^{SP}(n-1,1)
             +  4 \, I_{6}^{SP}(n)
             -  2 \, I_{12}^{SP}(0,n)
             +  2 \, I_{12}^{SP}(1,n-1)
             +  2 \, I_{13}^{SP}(0,n)
\nonumber \\
& & 
             -  2 \, I_{13}^{SP}(1,n-1)
             -  2 \, I_{14}^{SP}(0,n)
             +  2 \, I_{14}^{SP}(1,n-1)    \, \bigg \}
\nonumber \\
& & 
+ (\Delta\cdot p)^{-2} \bigg \{ \, 
                4 \, H_{6}^{SP}(n+1,0)
             -  8 \, H_{6}^{SP}(n,1)
             +  2 \, I_{1}^{SP}(0,n+1)
             -  2 \, I_{2}^{SP}(0,n+1)
\nonumber \\
& &
             +  4 \, I_{2}^{SP}(1,n)
             -  2 \, I_{3}^{SP}(n+1,0) 
             -  4 \, I_{4}^{SP}(n+1)
             -  4 \, I_{6}^{SP}(n+1)
            + (-1)^n 4 \, \widetilde I_{8}^{SP}(n-1,2) 
\nonumber \\
& & 
             +  2 \, I_{12}^{SP}(0,n+1)
             +  2 \, I_{12}^{SP}(2,n-1)
             -  4 \, I_{12}^{SP}(1,n)
             -  2 \, I_{13}^{SP}(0,n+1)
\nonumber \\
& &
             +  4 \, I_{13}^{SP}(1,n)
             -  2 \, I_{13}^{SP}(2,n-1)
             +  2 \, I_{14}^{SP}(0,n+1) 
             -  4 \, I_{14}^{SP}(1,n)
                                      \, \bigg \} \, \bigg ]
\nonumber \\
&=&  - \, \frac{g^4}{(4\pi)^4} \, (C_F^2 - C_F C_G /2) \,
            8 \, \frac{S_1(n)}{n} 
\ \ ,
\end{eqnarray}

\vspace{-0.3cm}
\begin{eqnarray}
\gamma_n^{(2j)} &=& + \, g^4 \, C_F C_G \,
           (\Delta\cdot p)^{-(n-1)} 
    \, \bigg [ \,
                    8 \, \hat H_{1}^{DP} (n)  
                  - 4 \, \hat H_{2}^{DP} (n) 
                  + 8 \, \hat H_3^{DP} (n) 
          +      I_{1}^{DP} (0,n-1) 
\nonumber \\
& &  
          +      I_{1}^{DP} (n-1,0) 
          - 4 \, I_{2}^{DP} (0,n-1) 
          - 5 \, I_{9}^{DP} (n-1,0)  
      + (\Delta\cdot p)^{-1} \bigg \{ \, 
            - 3 \, I_{1}^{DP}(0,n) 
\nonumber \\ 
& & 
            + 4 \, I_{1}^{DP}(1,n-1) 
            -      I_{1}^{DP}(n-1,1)   
            + 3 \, I_{2}^{DP}(0,n)   
            - 4 \, I_{2}^{DP}(1,n-1) 
            +      I_{2}^{DP}(n-1,1) 
                                      \, \bigg \} \, 
\nonumber \\
& & 
            - 8 \, \hat H_{2}^{SP} (n)
            + 2 \, I_{1}^{SP}(0,n-1) 
            + 2 \, I_{1}^{SP}(n-1,0) 
+ (\Delta\cdot p)^{-1} \bigg \{ \, 
               6 \, I_{1}^{SP}(0,n)    
\nonumber \\
& & 
             - 6 \, I_{1}^{SP}(n-1,1)  
             + 4 \, I_{1}^{SP}(1,n-1)      
             + 2 \, I_{2}^{SP}(0,n)    
             + 6 \, I_{2}^{SP}(n-1,1)    
\nonumber \\
& & 
             - 8 \, I_{2}^{SP}(1,n-1)  
             - 8 \, I_{3}^{SP}(n,0)   
             + 4 \, I_{3}^{SP}(n-1,1)    
             - 8 \, I_{6}^{SP}(n)
                                        \, \bigg \}    
\nonumber \\
& & 
+ (\Delta\cdot p)^{-2} \bigg \{ \,
             (-1)^{n+1} 16 \, \widetilde H_{6}^{SP}(n,1)  
             + 4 \, I_{1}^{SP}(n-1,2)
             - 8 \, I_{1}^{SP}(1,n)
             - 4 \, I_{2}^{SP}(n-1,2)
\nonumber \\
& & 
             + 8 \, I_{2}^{SP}(1,n)      
             - 8 \, I_{4}^{SP}(n+1)        
             + 8 \, I_{6}^{SP}(n+1) 
             - 8 \, I_{8}^{SP}(n-1,2)   
             +16 \, I_{8}^{SP}(n,1)  
                                        \, \bigg \} \, \bigg ]
\nonumber \\
&=&  0
\ \ ,
\end{eqnarray}
\vspace{-1.3cm}
\begin{eqnarray}
\gamma_n^{(2k)} &=& + \, g^4 \, (C_F^2 - C_F C_G /2) \,
          2 \, (\Delta\cdot p)^{-(n-1)} \, 
   \bigg [ \,    (-1)^{n} \, 8 \,  \hat H_4^{DP} (n)
          -      I_{11}^{DP} (n-1,0)  
          + 2 \, I_{12}^{DP} (n-1,0)  
\nonumber \\
& &
          +      I_{13}^{DP} (n-1,0)
          + 3 \, I_{14}^{DP} (n-1,0) 
      + (\Delta\cdot p)^{-1} \bigg \{ \, 
                   I_{2}^{DP}(n-1,1)
            -      I_{12}^{DP}(n,0) 
\nonumber \\
& & 
            +      I_{12}^{DP}(n-1,1) 
            +      I_{13}^{DP}(n,0)  
            -      I_{13}^{DP}(n-1,1)   
            -      I_{14}^{DP}(n-1,1) 
                                        \, \bigg \} \, 
\nonumber \\
& & 
            + 8 \, \{ 1 - (-1)^n \} \,  
                     \{ \hat H_4^{SP}(n) - \hat H_5^{SP}(n) \}
+ (\Delta\cdot p)^{-1} \bigg \{ \, 
              4 \, H_{7}^{SP} (n-1,1) 
            - 4 \, H_{7}^{SP} (n,0) 
\nonumber \\
& &
            - 4 \, \widetilde H_{7}^{SP} (n-1,1) 
            + 2 \, I_{2}^{SP} (n-1,1) 
            - 2 \, I_3^{SP}(n,0)
            - 2 \, \widetilde I_{3}^{SP} (n,0)   
            + 2 \, I_{12}^{SP} (n-1,1) 
\nonumber \\
& &
            - 2 \, I_{12}^{SP}(n,0)
            - 2 \, I_{13}^{SP} (n-1,1) 
            + 2 \, I_{13}^{SP}(n,0)
            - 2 \, I_{14}^{SP} (n-1,1) 
                                                   \, \bigg \} 
\nonumber \\
& &
+ (\Delta\cdot p)^{-2} \bigg \{ \, 
               (-1)^{n} \, 4 \, H_{6}^{SP} (n+1,0)
            + 8 \, H_{7}^{SP} (n,1) 
            - 4 \, H_{7}^{SP} (n+1,0) 
\nonumber \\
& &
            - 4 \, H_{7}^{SP} (n-1,2) 
            + 4 \, \widetilde H_{7}^{SP} (n-1,2) 
            + 2 \, I_3^{SP}(n+1,0)
            - 4 \, I_{8}^{SP} (n-1,2) 
            - 8 \, \widetilde I_{8}^{SP} (n,1) 
\nonumber \\
& &
            + 4 \, \widetilde I_{8}^{SP} (n-1,2) 
               \, \bigg \} \, \bigg ]
\nonumber \\
&=& - \, \frac{g^4}{(4\pi)^4} \, (C_F^2 - C_F C_G /2) \,
           8 \, \{ 1 - (-1)^n \} \, \frac{1}{n(n+1)}
\ \ ,
\end{eqnarray}
\begin{eqnarray}
\gamma_n^{(2n)} &=& + \, 4 \, g^4 \, C_F C_G \,
             (\Delta\cdot p)^{-(n-1)} \, \sum_{j=1}^{n-1} \, 
   \bigg [ \, 
                 I_{1}^{SP} (j,n-1-j) 
+ (\Delta\cdot p)^{-1} \bigg \{ \, 
          -      I_{1}^{SP} (n-1-j,j+1) 
\nonumber \\
& & 
          +      I_{1}^{SP} (j,n-j) 
          +      I_{2}^{SP} (n-1-j,j+1) 
          -      I_{2}^{SP} (j,n-j)   \, \bigg \} \, \bigg ]
\nonumber \\
&=& + \, \frac{g^4}{(4\pi)^4} \, C_F C_G \, 4 \, 
      \bigg [ \, \frac{1}{2n} \, \{ 4 S_1(n) - S_1^2(n) - S_2(n) \} 
               \, + \, S_3(n) \, - \, 2 \, \hat S(n) \, \bigg ]
\ \ ,
\end{eqnarray}

\vspace{-0.3cm}
\begin{eqnarray}
\gamma_n^{(2o)} &=& + \, g^4 \, (C_F^2 - C_F C_G /2) \,
          8 \, (\Delta\cdot p)^{-(n-1)} \, \sum_{j=1}^{n-1} \, 
   \bigg [ \, 
            -      I_{13}^{SP} (n-1-j,j) 
            +      I_{13}^{SP} (n-j,j-1) 
\nonumber \\
& &
            +      I_{14}^{SP}(n-1-j,j) 
            +      I_{14}^{SP}(n-j,j-1)  
      + (\Delta\cdot p)^{-1} \bigg \{ \, 
            +      I_{2}^{SP} (n-1-j,j+1) 
\nonumber \\
& &
            -      I_{2}^{SP} (n-j,j)   
            -      I_{12}^{SP} (n-1-j,j+1) 
            + 2 \, I_{12}^{SP} (n-j,j)  
            -      I_{12}^{SP} (n-j+1,j-1) 
\nonumber \\
& &
            +      I_{13}^{SP} (n-1-j,j+1) 
            - 2 \, I_{13}^{SP} (n-j,j)  
            +      I_{13}^{SP} (n-j+1,j-1) 
\nonumber \\
& &
            -      I_{14}^{SP} (n-1-j,j+1) 
            +      I_{14}^{SP} (n-j,j) 
                                            \, \bigg \} \, \bigg ]
\nonumber \\
&=& + \, \frac{g^4}{(4\pi)^4} \, (C_F^2 - C_F C_G /2) \, 8 \,
        \bigg [ \, 2 \, S_1 (n) \, 
           \left \{ \, \frac{1}{n} \, - 2 \, S_2'(n/2) \, 
                                      + 2 \, S_2(n)  \, \right \}
\nonumber \\
& &
\ \ \ \ \ \ \ \ \ \ \ \ \ \ \ \ \ \ 
\ \ \ \ \ \ \ \ \ \ \ \ \ \ \ \ \ \ 
           \, - \, 2 \, S_3(n) \, - \, S_3'(n/2) \, + \, 4 \, \hat S(n)
           \, + \, 8 \, \widetilde S (n) \, \bigg ]
\ \ .
\end{eqnarray}
In these equations, the color constants are 
$C_G=N_c$, $C_F=(N_c^2-1)/(2N_c)$, and $T_R=N_f / 2$, where
$N_c$ is the number of color and $N_f$ is the number of flavor.
The functions $S_1$, $S_2$, and others are defined by
$S_j(n)=\sum_{k=1}^n 1/k^j$, 
$S_j'(n/2)= [ \{ 1+(-1)^n \} S_j(n/2) + \{ 1-(-1)^n \} S_j((n-1)/2) ]/2$,
$\widetilde S(n)=\sum_{k=1}^n (-1)^k S_1(k) /k^2$,
and $\hat S(n)=\sum_{k=1}^n S_1(k) /k^2$.
We have to be careful in handling the $I_3(0,c)$, $I_8(0,c)$,
and $H_7(0,c)$ terms; however, the above results are valid also for $n=1$.
From the above expressions together with other diagram results
including the quark-field renormalization part in Refs. \cite{frs}
and \cite{gly}, we obtain the anomalous dimension for the structure
function $h_1$:
\begin{eqnarray}
\gamma_n^{total} &=& C_F^2 \, \bigg [ \, 32 \, S_1(n) \, 
             \{ \, S_2(n) \, - \, S_2'(n/2) \, \} 
            \, + \, 24 \, S_2(n) \, - \, 8 \, S_3'(n/2)
            \, + \, 64 \, \widetilde S(n) 
\nonumber \\
& &
\ \ \ \ \ \ \ \ 
            \, - \,  8 \, \frac{1-(-1)^n}{n(n+1)} \, - \, 3 \, \, \bigg ]
           + \, C_F \, T_R \, \bigg [ \, -\, \frac{160}{9} \, S_1(n)
         \, + \, \frac{32}{3} \, S_2(n) \, + \, \frac{4}{3} \, \, \bigg ]
\nonumber \\
& & 
           + \, C_F \, C_G \, \bigg [ \, \frac{536}{9} \, S_1(n)
             \, - \, \frac{88}{3} \, S_2(n) \, + \, 4 \, S_3'(n/2)
             \, - \, 32 \, \hat S(n) 
\nonumber \\
& & 
\ \ \ \ \ \ \ \ \ \ \ \ \ 
           - \, 16 \, S_1(n) \, \{ \,2 S_2(n) \, - \, S_2'(n/2) \, \}
         \, + \, 4 \, \frac{1-(-1)^n}{n(n+1)} \, - \, \frac{17}{3}
        \, \, \bigg ]
\ \ ,
\end{eqnarray}
where the factor $g^4/(4\pi)^4$ is taken out.

$\ \ \ $

\noindent
{\bf Appendix B: Integrals}

We list integrals $H_k$. The simple-pole terms are 
written with counter terms and the double-pole terms 
are without them. Finite contributions are neglected.
Because the $\hat H$ functions are used for $H_{1-5}$
in Appendix A, we express their results in the $\hat H$ form:

\begin{eqnarray}
H_1(a,b) &=& \int \frac{d^d k_1}{(2\pi)^d} \int \frac{d^d k_2}{(2\pi)^d}
              \, \frac{(\Delta\cdot k_1)^a 
              \,       (\Omega\cdot k_1)^b \, \ksl_1 }
           { (k_1^2)^2 \, (k_1-p)^2 \, (k_2-k_1)^2 \, (k_2-k_1+p)^2}
\ \ ,
\nonumber
\end{eqnarray}
\begin{eqnarray}
\hat H_1(a) &=&  + \frac{1}{(4\pi)^4}  \, (\Delta\cdot p)^{a-1} 
           \, \frac{1}{2a} \, \hat S_1(\epsilon,a)
\ \ ,
\label{int1}
\end{eqnarray}
where $\hat S_1(\epsilon,a)=-4/\epsilon^2-(2/\epsilon)/(a+1)$.
The following integrals are given by the above expression with
some replacements:

\noindent
$\hat H_2$: $(k_2-k_1)^2 \rightarrow k_2^2$ and
          $\hat S_1(\epsilon,a)=-2/\epsilon^2 
            +(1/\epsilon) [-1/(a+1)+1/a-2]$,

\noindent
$\hat H_3$: $(k_2-k_1+p)^2\rightarrow k_2^2$ and
          $\hat S_1(\epsilon,a)=-2/\epsilon^2 
            +(1/\epsilon) [S_1(a)-1/(a+1)-2]$,

\noindent
$\hat H_4$: $(k_1-p)^2 (k_2-k_1+p)^2 \rightarrow (k_2-p)^2 (p+k_1-k_2)^2$

          \ and $\hat S_1(\epsilon,a)=-(2/\epsilon^2)/a 
             +(1/\epsilon) [1/(a+1) - 1/a - 1/a^2]$,

\noindent
$\hat H_5$: $\ksl_1/[(k_1-p)^2 (k_2-k_1+p)^2]
               \rightarrow \ksl_2/[(k_2-p)^2 (p+k_1-k_2)^2]$.
          $\hat H_5 - \hat H_4$ is expressed by 

    \ Eq. (\ref{int1}) with $\hat S_1(\epsilon,a)=+(1/\epsilon)/a$.

\noindent
Different integral types are
\begin{eqnarray}
H_{6}(a,c)&=& \int \frac{d^d k_1}{(2\pi)^d} \int \frac{d^d k_2}{(2\pi)^d}
       \, \frac{(\Delta\cdot k_1)^a \, (\Delta\cdot k_2)^c \, k_2\cdot p}
           { (k_1^2)^2 \, (k_1-k_2)^2 \, k_2^2 \, (k_2-k_1+p)^2}
\nonumber \\
      &=& \frac{1}{(4\pi)^4} \, \frac{(-1)^c \, a! \, c!}{(a+c+1)! \, (a+c)}
           \, (\Delta\cdot p)^{a+c} \, 
            \bigg [ \, - \frac{1}{\epsilon^2}
                + \frac{1}{2\epsilon} \, 
             \bigg \{ -S_1(a+c)+S_1(a)
\nonumber \\
& & \ \ \ \ \ \ \ \ \ \ \ \ \ \ \ \ \ \ \ \ \ \ \ \ \ \ \ \ \ \ \ \ 
    \ \ \ \ \ \ \ \ \ \ \ \ \ \ \ \ \ \ 
       +S_1(c)  -\frac{1}{a+c+1} -\frac{1}{a} +1 \bigg \} \, \bigg ] 
\ \ ,
\label{int2}
\end{eqnarray}
\begin{eqnarray}
H_{7}(a,c)&=& \int \frac{d^d k_1}{(2\pi)^d} \int \frac{d^d k_2}{(2\pi)^d}
       \, \frac{(\Delta\cdot k_1)^a \, (\Delta\cdot k_2)^c \, k_1\cdot p}
           { (k_1^2)^2 \, (k_2-p)^2 \, (k_1-k_2)^2 \, (p+k_1-k_2)^2}
\nonumber \\  
&=& \frac{1}{(4\pi)^4} \, (\Delta\cdot p)^{a+c} \sum_{J=0}^c
     \frac{(a+c-J-1)! \, (J+1)! \, a! \, c!}
          {(c-J)! \, (a+c)! \, (a+J+1)!} \, \bigg [ - \frac{1}{\epsilon^2} 
\nonumber \\
& &               + \frac{1}{2\epsilon} 
        \bigg \{ -2 S_1(a+c+1) + S_1(a+c-J-1) + S_1(a+J+1) 
        \bigg \} \, \bigg ]
\ \ .
\label{int4}
\end{eqnarray}

\vfill\eject

$\ \ \ $

\vspace{-0.8cm}

\vfill\eject

$\ \ \ $

\vspace{1.0cm}

\vspace{2.0cm}

\end{document}